\documentclass[11pt]{article}
\usepackage{graphicx}

\newcommand{\Nf}{N_{\!f}}
\newcommand{\MSbar}{\overline{\mbox{MS}}}

\newcommand{\text}{\rm}
\newcommand{\al}{\alpha}
\newcommand{\p}{\partial}
\newcommand{\oc}{\overline{c}}
\newcommand{\F}{\vphantom{x}_2F_1}
\newcommand{\G}{\vphantom{x}_3F_2}
\newcommand{\HPQ}{\vphantom{x}_pF_q}
\newcommand{\s}{\sigma}
\newcommand{\omu}{\overline{\mu}}
\newcommand{\lms}{\Lambda_{\overline{\mbox{\tiny{MS}}}}}
\newcommand{\Evac}{E_{\textrm{\tiny{vac}}}}
\newcommand{\mf}{\mathcal{F}}
\newcommand{\wj}{\widetilde{J}}
\newcommand{\wsigma}{\widetilde{\sigma}}
\newcommand{\mw}{\mathcal{W}}

\begin{document}

\title{\textbf{Dynamical gluon mass generation from \\}%
\textbf{\ $\left\langle A_{\mu }^{2}\right\rangle$ in linear
covariant gauges }}
\author{D. Dudal\thanks{%
Research Assistant of The Fund For Scientific Research-Flanders,
Belgium.} \
and H. Verschelde\thanks{%
david.dudal@ugent.be, henri.verschelde@ugent.be} \\
%EndAName
{\small {\textit{Ghent University }}}\\
{\small {\textit{Department of Mathematical Physics and Astronomy,
Krijgslaan 281-S9, }}}\\{\small {\textit{B-9000 Gent,
Belgium}}}\and
J.A. Gracey\thanks{%
jag@amtp.liv.ac.uk } \\
%EndAName
{\small {\textit{Theoretical Physics Division}}} \\
{\small {\textit{Department of Mathematical Sciences}}} \\
{\small {\textit{University of Liverpool }}} \\
{\small {\textit{P.O. Box 147, Liverpool, L69 3BX, United
Kingdom}}} \and V.E.R. Lemes, M.S. Sarandy,
R.F. Sobreiro, and S.P. Sorella\thanks{%
vitor@dft.if.uerj.br, sarandy@dft.if.uerj.br,
sobreiro@dft.if.uerj.br,
sorella@uerj.br} \\
%EndAName
{\small {\textit{UERJ - Universidade do Estado do Rio de Janeiro,}}} \\
{\small {\textit{\ Rua S\~{a}o Francisco Xavier 524, 20550-013
Maracan\~{a}, }}} {\small {\textit{Rio de Janeiro, Brazil}}}  }
\date{}
\maketitle

\vspace{-12cm} \hfill LTH--611 \vspace{12cm}

%\newpage

\begin{abstract}
We construct the multiplicatively renormalizable effective
potential for the mass dimension two local composite operator
$A_\mu^a A^{\mu a}$ in linear covariant gauges. We show that the
formation of $\left\langle A_\mu^a A^{\mu a}\right\rangle$ is
energetically favoured and that the gluons acquire a dynamical
mass due to this gluon condensate. We also discuss the gauge
parameter independence of the resultant vacuum energy.
\end{abstract}

\newpage

\section{Introduction.}
In a previous paper \cite{Dudal:2003np}, we took the first step
towards constructing a renormalizable effective potential for the
local composite operator (LCO) $A_\mu^2\equiv A_\mu^a A^{\mu a}$
in linear covariant gauges. It was shown within the algebraic
renormalization formalism \cite{book,Barnich:2000zw} that
$A_\mu^2$ is multiplicatively renormalizable to all orders in
perturbation theory. At the same time, the anomalous dimension of
$A_\mu^2$ was explicitly computed to 2-loops in the $\MSbar$
scheme as a function of the gauge fixing parameter, $\al$, where
$\al=0$ corresponds to the Landau gauge. The computation exploited
the fact that in linear covariant gauges the operator does not
mix, for example, with ghost operators of dimension two with the
same quantum numbers.

\noindent The operator $A_\mu^2$ has recently received widespread
interest in Yang-Mills theory in the Landau gauge. Its relevance
has been advocated both from a theoretical point of view as well
as from lattice simulations
\cite{Gubarev:2000eu,Gubarev:2000nz,Boucaud:2000nd,Boucaud:2001st,
Boucaud:2002nc}. Analytic results in favour of a non-zero value
for the condensate $\left\langle A_\mu^2\right\rangle$ in the
Landau gauge have been obtained recently,
\cite{Verschelde:2001ia,Dudal:2003vv}. Further, the inclusion of
quarks has been considered in \cite{Browne:2003uv}. Motivated by
the result of \cite{Gracey:2002yt} it has been shown in
\cite{Dudal:2002pq} that $A_\mu^2$ is multiplicatively
renormalizable to all orders in the Landau gauge, but its
anomalous dimension is given by a combination of the gauge beta
function, $\beta(a)$, and the anomalous dimension, $\gamma_A(a)$,
of the gluon field, according to the relation
\cite{Gracey:2002yt,Dudal:2002pq}
\begin{equation}\label{1}
\gamma_{A^2}(a) ~=~ -~ \left(\frac{\beta(a)}{a} ~+~
\gamma_A(a)\right), \hspace{2cm}a~=~\frac{g^2}{16\pi^2}\;.
\end{equation}
An important consequence of the formation of the $\left\langle
A_\mu^2\right\rangle$ condensate in the Landau gauge is the
dynamical generation of a gluon mass
$m_{\mbox{\tiny{\textrm{gluon}}}}\approx 500$MeV
\cite{Verschelde:2001ia}. Lattice simulations of the $SU(2)$ gluon
propagator in the Landau gauge report
$m_{\mbox{\tiny{\textrm{gluon}}}}\approx 600$MeV,
\cite{Langfeld:2001cz}. Gluon masses have also been extracted from
lattice methods in the Laplacian
\cite{Alexandrou:2001fh,Alexandrou:2002gs} and Maximal Abelian
gauges \cite{Amemiya:1998jz,Bornyakov:2003ee}. Earlier the pairing
of gluons was discussed in connection with mass generation as a
result of the instability of the perturbative Yang-Mills vacuum
\cite{Fukuda:1977mz,Fukuda:1977wj,Savvidy:1977as,Gusynin:1978tr,
Cornwall:1981zr}. A dynamical gluon mass might also be important,
for example, in connection with the glueball spectra,
\cite{Cornwall:1982zn,Greensite:1985vq}. A dimension two gluon
condensate $\left\langle A_i^2\right\rangle$ was already
introduced in \cite{Greensite:1985vq}, where the Coulomb gauge was
considered. Furthermore, a dynamical gluon mass is part of a
certain criterion for confinement introduced by Kugo and Ojima,
\cite{Kugo:gm,ko1}. For a recent review see \cite{Alkofer:2000wg}.

\noindent It is no coincidence that the Landau gauge is employed
in the search for a gluon condensate of mass dimension two. As is
well known, there does not exist a local, gauge-invariant operator
of mass dimension two in Yang-Mills theories. However, a
\emph{non-local} gauge invariant dimension two operator can be
constructed by minimizing $A_\mu^2$ along each gauge orbit
\cite{semenov,Zwanziger:tn,Stodolsky:2002st}, $A^2_{\min}\equiv
(VT)^{-1}\min_{U}\int d^{4}x\left(A_{\mu}^U\right)^2$ where $VT$
is the space time volume and $U$ is a generic $SU(N)$
transformation. This operator $A^{2}_{\min}$ is related to the
so-called fundamental modular region (FMR), which is the set of
absolute minima of $(VT)^{-1}\int d^{4}x\left(
A_{\mu}^U\right)^2$. In the Landau gauge, $\p_\mu A^{a\mu}=0$, so
that $A^2_{\min}$ and $A_\mu^2$ coincide within the FMR. As such,
a gauge invariant meaning can indeed be attached to $\left\langle
A_\mu^2\right\rangle$ in the Landau gauge, as it was also
expressed in \cite{Verschelde:2001ia}.

\noindent Another interesting property of the Landau gauge is that
the operator $A_\mu^2$ is BRST invariant on-shell. If one
considered alternative gauges to the Landau gauge, one could
search for a class of gauges in which the operator $A_\mu^2$ can
be generalized to a mass dimension two operator while maintaining
the on-shell BRST invariance. Doing so, one should consider a
class of non-linear covariant gauges, which are the so-called
Curci-Ferrari gauges \cite{Curci:bt,Curci:1976ar}, where $A_\mu^2$
is generalized to the mixed gluon-ghost operator
$\left(\frac{1}{2}A_\mu^a A^{\mu a}+\alpha \oc^a c^a\right)$
\cite{Kondo:2001nq,Kondo:2001tm}. The latter operator is indeed
BRST invariant on-shell \cite{Kondo:2001nq,Kondo:2001tm}, and has
been proven to be multiplicatively renormalizable to all orders
\cite{Dudal:2003pe}, and to give rise to a dynamical gluon mass in
the Curci-Ferrari gauge \cite{Dudal:2003gu}. Moreover, in
\cite{Kondo:2003uq}, the physical meaning of
$\left(\frac{1}{2}A_\mu^a A^{\mu a}+\alpha \oc^a c^a\right)$ was
discussed, based on on-shell BRST invariance.

\noindent In the Maximal Abelian gauge, which is a renormalizable
gauge in the continuum \cite{Min:bx,Fazio:2001rm}, one should
consider the operator
$\left(\frac{1}{2}A_\mu^{\beta}A^{\mu\beta}+\xi\oc^{\beta}c^{\beta}\right)$
where the group index $\beta$ labels the off-diagonal generators
of $SU(N)$ with $\beta=1,\ldots,N(N-1)$ and $\xi$ is the gauge
parameter of the Maximal Abelian gauge. This operator also enjoys
the property of being both BRST invariant on-shell
\cite{Kondo:2001nq,Kondo:2001tm} and multiplicatively
renormalizable to all orders \cite{Dudal:2003pe,Ellwanger:2002sj}.
Although the effective potential for the condensate
$\left\langle\frac{1}{2}A_\mu^{\beta}A^{\mu\beta}+\xi\oc^{\beta}c^{\beta}
\right\rangle$ has not yet been obtained, we expect it to have a
non-vanishing vacuum expectation value, which would result in a
dynamical mass for the off-diagonal gluons. This is based on the
close similarity between the Maximal Abelian gauge and the
Curci-Ferrari gauge and hence the results of \cite{Dudal:2003gu}.

\noindent More commonly, the Landau gauge is a special case of the
well known linear covariant gauges. Although the operator
$A_\mu^2$ is not even BRST invariant on-shell in these gauges, it
is still renormalizable to any order in perturbation theory
\cite{Dudal:2003np}. This is due to the fact that, thanks to the
additional shift symmetry, $\oc\rightarrow\oc+\textrm{const}$, of
the antighost in the linear covariant gauges, the composite
operator $A_\mu^2$ does not mix into the dimension two ghost
operator $\oc c$. In this article, we will construct the effective
potential for the dimension two condensate $\left\langle
A_\mu^2\right\rangle$ in linear gauges and show that a
non-vanishing value of $\left\langle A_\mu^2\right\rangle$ is
energetically favourable, resulting in dynamical gluon mass
generation.

\noindent The paper is organized as follows. In section 2, we
briefly review the local composite operators formalism and
explicitly calculate the 1-loop effective potential. In section 3,
we discuss the gauge parameter independence of the vacuum energy
which requires an extension of the LCO formalism. The behaviour of
the gluon propagator is discussed briefly in section 4 whilst we
provide concluding comments in section 5.

\section{LCO formalism and effective potential for $A_\mu^2$.}
\subsection{Construction of a renormalizable effective action for
$A_\mu^2$.} We begin with the Yang-Mills action in linear
covariant gauges
\begin{eqnarray}
S &=&S_{YM}+S_{GF+FP}  \label{sym} \\
&=&-\frac{1}{4}\int d^{4}xF_{\mu \nu }^{a}F^{a\mu \nu }+\int
d^{4}x\left(
b^{a}\partial _{\mu }A^{a\mu }+\frac{\alpha }{2}b^{a}b^{a}+\overline{c}%
^{a}\partial ^{\mu }D_{\mu }^{ab}c^{b}\right) \;,  \nonumber
\end{eqnarray}
where
\begin{equation}
D_{\mu }^{ab}\equiv \partial _{\mu }\delta ^{ab}-gf^{abc}A_{\mu
}^{c}\;, \label{cd}
\end{equation}
is the covariant derivate in the adjoint representation. In order
to study the local composite operator $A_\mu^2$, we introduce it
into the action by means of a BRST\ doublet \cite{book} of
external sources $\left( J,\lambda \right) $, namely
\begin{equation}
S_{J}=s\int d^{4}x\left( \frac{1}{2}\lambda A_{\mu }^2+\frac{\zeta
}{2}\lambda J\right) =\int d^{4}x\left( \frac{1}{2}JA_{\mu
}^2+\lambda A^{a}_\mu\partial ^{\mu }c^{a}+\frac{\zeta
}{2}J^{2}\right) \;, \label{sj}
\end{equation}
where $s$ denotes the BRST\ nilpotent operator acting as
\begin{eqnarray}
sA_{\mu }^{a} &=&-D_{\mu }^{ab}c^{b}\;,  \nonumber \\
sc^{a} &=&\frac{1}{2}gf^{abc}c^{b}c^{c}\;,  \nonumber \\
s\overline{c}^{a} &=&b^{a}\;,  \nonumber \\
sb^{a} &=&0\;,  \nonumber \\
s\lambda &=&J\;,  \nonumber \\
sJ &=&0\;.  \label{s}
\end{eqnarray}
According to the local composite operator technique
\cite{Verschelde:2001ia,Verschelde:jj,Verschelde:jx,Knecht:2001cc},
the dimensionless parameter $\zeta$ is needed to account for the
divergences present in the vacuum Green function $\left\langle
A_\mu^{2}(x)A_\nu^{2}(y)\right\rangle $, which turn out to be
proportional to $J^{2}$. As is apparent from the expressions
(\ref{sym}) and
(\ref{sj}), the action $%
\left( S_{YM}+S_{GF+FP}+S_{J}\right) $ is BRST invariant
\begin{equation}
s\left( S_{YM}+S_{GF+FP}+S_{J}\right) =0\;.  \label{si}
\end{equation}
As was shown in \cite{Dudal:2003np}, the action $\left(
S_{YM}+S_{GF+FP}+S_{J}\right)$ enjoys the property of being
multiplicatively renormalizable to all orders of perturbation
theory.

\noindent To obtain the effective potential, we set the source
$\lambda$ to zero and consider the renormalized generating
functional
\begin{equation}\label{d1}
    \exp-i\mw(J)=\int [D\varphi]\exp iS(J)\;,
\end{equation}
with
\begin{equation}\label{d2}
    S(J)=S_{YM}+S_{GF+FP}+S_{CT}+\int d^{4}x
\left(Z_2J\frac{A_\mu^2}{2}+(\zeta+\delta\zeta)\frac{J^2}{2}\right)\;,
\end{equation}
where $\varphi$ denotes the relevant fields and $S_{CT}$ is the
usual counterterm contribution. Also, $\delta\zeta$ is the
counterterm accounting for the divergences proportional to $J^2$.
The bare quantities are given by \cite{Dudal:2003np}
\begin{eqnarray}
A^{\mu a}_o&=&Z_A^{1/2}A^{\mu
a}\;,\hspace{1cm}c_o^a=Z_c^{1/2}c^a\;,
\hspace{1cm}\oc_o^a=Z_c^{1/2}\oc^a\;,\hspace{1cm}b_o^a=Z_A^{-1/2}b^a\;,\nonumber\\
g_o&=&Z_g g\;,\hspace{1.7cm}\al_o=Z_A
\al\;,\hspace{1.35cm}\zeta_o=Z_\zeta
\zeta\;,\hspace{1.45cm}J_o=Z_J J\;,\nonumber\\
\end{eqnarray}
where $Z_\zeta\zeta=\zeta+\delta\zeta$ and $Z_J=\frac{Z_2}{Z_A}$.
The functional $\mw(J)$ obeys the renormalization group equation
(RGE)
\begin{equation}\label{rge3}
    \left(\mu\frac{\p}{\p\mu}+\beta(g^2)\frac{\p}{\p
g^2}+\al\gamma_\al(g^2)\frac{\p}{\p\al}-\gamma_{A^2}(g^2)\int d^4x
J\frac{\delta}{\delta
    J}+\eta(g^2,\zeta)\frac{\p}{\p\zeta}\right)\mw(J)=0\;,
\end{equation}
where
\begin{eqnarray}\label{rge4}
    \beta(g^2)&=&\mu\frac{\p}{\p\mu}g^2\;,\nonumber\\
\gamma_{\al}(g^2)&=&\mu\frac{\p}{\p\mu}\ln\al=\mu\frac{\p}{\p\mu}\ln
    Z_A^{-1}=-2\gamma_A(g^2)\;,\nonumber\\
\gamma_{A^2}(g^2)&=&\mu\frac{\p}{\p\mu}\ln Z_J\;,\nonumber\\
\eta(g^2,\zeta)&=&\mu\frac{\p}{\p\mu}\zeta\;.
\end{eqnarray}
{}From the bare Lagrangian, we infer that
\begin{equation}\label{rge5}
    \zeta_oJ_o^2=\mu^{-\varepsilon}(\zeta+\delta\zeta)J^2\;,
\end{equation}
where we will use dimensional regularization throughout with the
convention that $d=4-\varepsilon$. Hence
\begin{equation}\label{rge6}
\mu\frac{\p}{\p\mu}\zeta=\eta(g^2,\zeta)=2\gamma_{A^2}(g^2)\zeta+\delta
(g^2,\al)\;,
\end{equation}
with
\begin{equation}  \label{rge7}
\delta(g^{2},\alpha)=\left(\varepsilon+2\gamma_{A^2}(g^{2},\alpha)-\beta
(g^{2})\frac{\partial }{\partial g^{2}}%
-\alpha\gamma_{\alpha}(g^{2},\alpha)\frac{\partial}{\partial\alpha}%
\right)\delta\zeta\;.
\end{equation}
Now, we are faced with the problem of the hitherto arbitrary
parameter $\zeta$. As explained in
\cite{Verschelde:2001ia,Verschelde:jj,Verschelde:jx,Knecht:2001cc},
setting $\zeta=0$ would give rise to an inhomogeneous RGE for
$\mw(J)$
\begin{equation}  \label{rge8}
    \left(\mu\frac{\p}{\p\mu}+\beta(g^2)\frac{\p}{\p
g^2}+\al\gamma_\al(g^2)\frac{\p}{\p\al}-\gamma_{A^2}(g^2)\int d^4x
J\frac{\delta}{\delta
    J}\right)\mw(J)=\delta(g^2,\al)\int d^4x \frac{J^2}{2}\;,
\end{equation}
and a non-linear RGE for the effective action $\Gamma$ for the
composite operator $A_\mu^2$. This problem can be overcome by
making $\zeta$ a function of $g^2$ and $\al$ so that, if $g^2$
runs according to $\beta(g^2)$ and $\al$ according to
$\gamma_\al(g^2)$, $\zeta(g^2,\al)$ will run according to
(\ref{rge6}). This is accomplished by setting $\zeta$ equal to the
solution of the differential equation
\begin{equation}\label{rge9}
\left(\beta(g^2)\frac{\partial}{\partial
g^2}+\al\gamma_\al(g^2,\al)\frac{\partial}{\partial\al}\right)\zeta(g^2,
\al)=2\gamma_{A^2}(g^2)\zeta(g^2,\alpha)+\delta(g^2,\al)\;.
\end{equation}
Since $\zeta(g^2,\al)$ now automatically runs according to its
RGE, $\mw(J)$ obeys the homogeneous renormalization group equation
\begin{equation}  \label{rge10}
    \left(\mu\frac{\p}{\p\mu}+\beta(g^2)\frac{\p}{\p
g^2}+\al\gamma_\al(g^2)\frac{\p}{\p\al}-\gamma_{A^2}(g^2)\int d^4x
J\frac{\delta}{\delta J}\right)\mw(J)=0\;.
\end{equation}
The final step in the formal construction of the effective
potential for $\left\langle A_\mu^2\right\rangle$ is the removal
of the $J^2$ terms from the Lagrangian by means of a renormalized
Hubbard-Stratonovich transformation. By this procedure, the energy
interpretation of the effective action is made explicit again and
the conventional 1PI machinery applies. We insert unity written as
\begin{equation}  \label{rge11}
1=\frac{1}{N}\int [D\sigma]\exp\left[i\int
d^{4}x\left(-\frac{1}{2Z_\zeta\zeta}\left(%
\frac{\sigma}{g}-Z_2\frac{A_\mu^2}{2}-Z_\zeta\zeta
J\right)^{2}\right)\right]\;,
\end{equation}
with $N$ the appropriate normalization factor, in (\ref{d1}) to
arrive at the Lagrangian
\begin{eqnarray}
\mathcal{L}(A_\mu,\sigma)=-\frac{1}{4}F_{\mu\nu}^{2}+\mathcal{L}_{GF+FP
}+\mathcal{L}_{CT}-\frac{\sigma^2}{2g^2Z_\zeta\zeta}+\frac{1}{2}\frac{Z_2
}{g^2Z_\zeta\zeta}g\sigma
A_{\mu}^2-\frac{1}{8}\frac{Z_2^2}{Z_\zeta\zeta}\left(A_\mu^2\right)^2\;,
\label{rge12}
\end{eqnarray}
while
\begin{eqnarray}
\label{rge13}\exp -i\mw(J)&=&\int [D\varphi]\exp iS_\sigma(J)\;,\\
\label{rge13bis}S_\sigma(J)&=&\int
d^4x\left(\mathcal{L}(A_\mu,\sigma)+J\frac{\sigma}{g}\right)\;.
\end{eqnarray}
Now, the source $J$ appears as a linear source term for
$\frac{\sigma}{g}$. From (\ref{d1}) and (\ref{rge13}), one has the
following identification
\begin{equation}\label{rge14}
    \left.\frac{\delta\mw(J)}{\delta
J}\right|_{J=0}=-\left\langle\frac{A_\mu^2}{2}\right\rangle=-\left\langle
\frac{\sigma}{g}\right\rangle\;,
\end{equation}
where we will not write the renormalization factors from now on.
This equation states that the gauge condensate $\left\langle
A_\mu^2\right\rangle$ is related to the expectation value of the
field $\s$, evaluated with the new action, $\int
d^{4}x\mathcal{L}(A_\mu,\sigma)$, of (\ref{rge12}).

\noindent Although we have not considered the contribution from
(massless) quark fields in the previous analysis, it can be
checked that the results remain unchanged if matter fields are
included.

\subsection{Explicit calculation of the 1-loop effective potential.}
Firstly, we will determine the renormalization group function
$\delta(g^2,\al)$ as defined in (\ref{rge7}). All the following
results will be within the $\MSbar$ scheme. The value for
$\beta(g^2)$ can be found in the literature. In $d$ dimensions,
one has
\begin{eqnarray}\label{rge15}
    \beta(g^2)&=&-\varepsilon g^2-2\left(\beta_0 g^4+\beta_1
g^6+O(g^8)\right)\;,\nonumber\\
\beta_0&=&\frac{1}{16\pi^2}\left(\frac{11}{3}C_A-\frac{4}{3}T_FN_f\right)
\;,\nonumber\\
\beta_1&=&\frac{1}{\left(16\pi^2\right)^2}\left(\frac{34}{3}C_A^2-4C_FT_FN_f
-\frac{20}{3}C_AT_FN_f\right)\;.
\end{eqnarray}
where the Casimirs of the colour group are defined by
$\textrm{Tr}(T^a T^b)$~$=$~$T_F\delta^{ab}$, $T^aT^a$~$=$~$C_F I$,
$f^{acd}f^{bcd}$~$=$~$C_A\delta^{ab}$ and $\Nf$ is the number of
quark flavours. For $\gamma_\al(g^2)$, we use the relation
$\gamma_\al(g^2)=-2\gamma_A(g^2)$. The anomalous dimension
$\gamma_A(g^2)$ of the gluon field in linear covariant gauges was
calculated at three loops in $\MSbar$ in \cite{Larin:tp}. Adapting
that result to our convention, the anomalous dimension of the
gauge parameter is
\begin{eqnarray}\label{rge16}
    \gamma_\al(g^2)&=&a_0g^2+a_1g^4+O(g^6)\;,\nonumber\\
a_0&=&\frac{1}{16\pi^2}\left(C_A\left(\frac{13}{3}-\al\right)-\frac{8}{3}
T_F \Nf\right)\;,\nonumber\\
a_1&=&\frac{1}{\left(16\pi^2\right)^2}\left(C_A^2\left(\frac{59}{4}
-\frac{11}{4}\al-\frac{1}{2}\al^2\right)
    -10C_AN_fT_F-8C_FN_fT_F\right)\;.
\end{eqnarray}
The anomalous dimension, $\gamma_{A^2}(g^2)$, of the composite
operator $A_\mu^2$ was calculated in \cite{Dudal:2003np} and reads
\begin{eqnarray}\label{rge17}
\gamma_{A^2}(g^2)&=&\gamma_0g^2+\gamma_1 g^4+O(g^6)\;,\nonumber\\
\gamma_0&=&\frac{1}{6}\frac{1}{16\pi^2}\left[ \left( 35+3\alpha
\right) C_{A}-16T_{F}N_{\!f}\right]\;,\nonumber\\
\gamma_1&=&\frac{1}{24}\frac{1}{\left(16\pi^2\right)^2}\left[
\left( 449+33\alpha +18\alpha ^{2}\right)
C_{A}^{2}-280C_{A}T_{F}N_{\!f}-192C_{F}T_{F}N_{\!f}\right]\;.
\end{eqnarray}
In order to determine $\delta(g^2,\al)$, we still require the
counterterm $\delta\zeta$. In principle, this can be directly
calculated from the divergences in $\mw(J)$ when the propagator
for a gluon with mass $J$ is used. However, a less cumbersome way
to compute $\delta\zeta$ was described in \cite{Browne:2003uv}. It
is based on the fact that the divergences arise in the $O(J^2)$
term and therefore that part of the Green's function which
contains these divergences is equivalent to the Green's function
with a double insertion of the $J A_\mu^2$ operator. More
specifically, one has two external $J$ insertions with a non-zero
momentum flowing into one insertion where the only internal
couplings are those of the usual QCD action. Moreover, one does
not require massive propagators but instead can use massless
fields which simplifies the calculation. Therefore one is reduced
to computing a massless two-point function for which the {\sc
Mincer} algorithm, \cite{Gorishnii:1989gt}, was designed. We used
the version written in {\sc Form},
\cite{Larin:1991fz,Vermaseren:2000nd}, where the Feynman diagrams
are generated by {\sc Qgraf}, \cite{Nogueira:1993ex}, to determine
the divergence structure to three loops. Although we only require
the result to two loops the extra loop evaluation in fact acts as
a non-trivial check on the two loop result. This is because the
emergence of the correct double and triple poles in $\varepsilon$
at three loops, in a way which is consistent with the
renormalization group, verifies that the single and double poles
of the two loop expression for $\delta \zeta$ are correct. We
found
\begin{eqnarray}\label{rge18}
\delta\zeta&=&\frac{2}{\varepsilon}\frac{N_A}{16\pi^2}\left(-\frac{3}{2}
-\frac{\alpha^2}{2}\right)+\frac{N_Ag^2}{\left(16\pi^2\right)^2}\left[
\frac{4}{\varepsilon^2}\left(C_A\left(\frac{35}{8}+\frac{3}{8}\al+\frac{3}
{ 8}\al^2+\frac{3}{8}\al^3\right)
-2T_FN_f\right)\right.\nonumber\\&+&\left.\frac{2}{\varepsilon}
\left(C_A\left(-\frac{139}{12}-\frac{5}{8}\al-\frac{1}{2}\al^2-\frac{1}{8}
\al^3\right)
    +\frac{8}{3}T_FN_f\right)\right]+O(g^4)\;,
\end{eqnarray}
where $N_A$ is the dimension of the adjoint representation of the
colour group. Assemblying our results leads to
\begin{eqnarray}\label{rge19}
\delta(g^2,\al)\nonumber&=&\delta_0+\delta_1g^2+O(g^4)\;,\nonumber\\
    \delta_0&=&\frac{N_A}{16\pi^2}\left(-3-\al^2\right)\;,\nonumber\\
\delta_1&=&\frac{1}{6}\frac{N_A}{\left(16\pi^2\right)^2}\left(C_A\left(
-278-15\al -12\al^2-3\al^3\right)+64T_FN_f\right)\;.
\end{eqnarray}
As a check we see that $\delta(g^2,\al)$ contains no poles for
$\varepsilon\rightarrow 0$. Further, the expressions (\ref{rge19})
lead to the same results which were obtained earlier in the case
of the Landau gauge ($\al=0$), as can be inferred from
\cite{Verschelde:2001ia} without quarks and \cite{Browne:2003uv}
with quarks.

\noindent {}From the renormalization group functions
(\ref{rge15}), (\ref{rge16}), (\ref{rge17}) and (\ref{rge19}), it
is easy to see that the equation (\ref{rge9}) can be solved for by
expanding $\zeta(g^2,\al)$ in a Laurent series as
\begin{equation}\label{rge20}
    \zeta(g^2,\al)=\frac{\zeta_0(\al)}{g^2}+\zeta_1(\al)+O(g^2)\;.
\end{equation}
Substituting this expression in equation (\ref{rge9}), we obtain
\begin{eqnarray}
\label{rge21a}2\beta_0\zeta_0+\al
a_0\frac{\partial\zeta_0}{\partial\al}&=&2\gamma_0\zeta_0+\delta_0\;,\\
\label{rge21b}2\beta_1\zeta_0+\al
a_0\frac{\partial\zeta_1}{\partial\al}+\al
a_1\frac{\partial\zeta_0}{\partial\al}&=&2\gamma_0\zeta_1+2\gamma_1\zeta_0
+\delta_1\;.
\end{eqnarray}
Thus (\ref{rge21a}) gives
\begin{eqnarray}\label{rge22}
\zeta_0(\al)&=&\frac{2\al
C_0+3\left(78-26\al^2+3\al^3+18\al\ln\left|\al\right|\right)C_AN_A+48\left(\al^2-3\right)
N_A \Nf T_F}{2\left((3\al-13)C_A+8N_fT_F\right)^2}\;,
\end{eqnarray}
with $C_0$ a constant of integration. As a consequence of the
already rather complicated structure of $\zeta_0$, we will
determine $\zeta_1$ without quarks present corresponding to
$\Nf=0$ since the expression for $\zeta_1$ with $\Nf\neq0$ is
several pages long. Using \verb"Mathematica", we find
\begin{eqnarray}\label{rge24}
\zeta_1(\al)&=&-\frac{1}{1220736\pi^2(13-3\al)^4}\left(-1220736\pi^2
\al^{35/13}\left|-13+3\al\right|^{4/13}C_1
\right.\nonumber\\&+&\left.12716C_0\al^2\left(-442-132\al+54\al^2-1287\left(
1-\frac{3\al}{13}\right)^{4/13}\alpha
\F\left[\frac{4}{13},\frac{4}{13};\frac{17}{13};\frac{3\al}{13}\right]\right)
C_A\right.\nonumber\\
&+&\left.\left(1697175909\left(1-\frac{3\al}{13}\right)^{4/13}\al^3\G\left
[\frac{4}{13},\frac{4}{13},\frac{4}{13};\frac{17}{13},\frac{17}{13};
\frac{3\al}{13}\right]\right.
\right.\nonumber\\&+&\left.\left.3335904\left(1-\frac{3\al}{13}\right)^
{4/13}\al^4\G\left[\frac{17}{13},\frac{17}{13},\frac{17}{13};\frac{30}{13},
\frac{30}{13};\frac{3\al}{13}\right]\right.\right.\nonumber\\
&+&\left.\left.17\left(-396870474+368850105\al-48761440\al^2+2066214\al^3
+1928718\al^4 \right.\right.\right.\nonumber \\
&-& \left.\left.\left. 1004751\al^5
+60588\al^6-12894024\left(1-\frac{3\al}{13}\right)^{4/13}
\al^2\F\left[-\frac{9}{13},-\frac{9}{13};\frac{4}{13};\frac{3\al}{13}
\right]
\right.\right.\right.\nonumber\\&-&\left.\left.\left.833976\left(1-\frac{3\al}
{13}\right)^{4/13}\al^4\F\left[\frac{4}{13},\frac{17}{13};\frac{30}{13};
\frac{3\al}{13}\right]
-8926632\al^2\ln\left|\al\right|\right.\right.\right.\nonumber\\&+&\left.
\left.\left.
2059992\al^3\ln\left|\al\right|+833976\left(1-\frac{3\al}{13}\right)^{4/13}
\al^4\F\left[\frac{4}{13},\frac{17}{13};\frac{30}{13};\frac{3\al}{13}\right]
\ln\left|\al\right|\right.\right.\right.
\nonumber\\&-&\left.\left.\left.43758\al^3
\left(-1961+702\ln\left|\al\right|\right)\right)
\right)N_A\right)\;,
\end{eqnarray}
where $C_1$ is a constant of integration and the (generalized)
hypergeometric function is
\begin{eqnarray}\label{rge23}
\HPQ\left[a_1,\cdots,a_p;b_1,\cdots,b_q;z\right]=\sum_{k=0}^{+\infty}
\frac{\left(a_1\right)_k\cdots\left(a_p\right)_k}{\left(b_1\right)_k\cdots
\left(b_q\right)_k}\frac{z^k}{k!}\;,
\end{eqnarray}
where
\begin{equation}\label{rge23bis}
(a)_k=a(a+1)\cdots(a+k-1)\;,
\end{equation}
is the Pochhammer symbol. We note that
$\zeta_0(\al=0)=\frac{9N_A}{13C_A}$ and
$\zeta_1(\al=0)=\frac{161}{832}\frac{N_A}{\pi^2}$, which recovers
the Landau gauge results of
\cite{Verschelde:2001ia,Browne:2003uv}. Further, the constants of
integration $C_0$ and $C_1$ do no enter the Landau gauge results.

\noindent {}From expression (\ref{rge12}), we deduce that the tree
level gluon mass is provided by
\begin{equation}\label{rge25}
    m^2=\frac{g\sigma}{\zeta_0}\;,
\end{equation}
while the 1-loop effective potential becomes
\begin{eqnarray}\label{rge26}
V_1(\sigma)&=&\frac{\s^2}{2\zeta_0}\left(1-\frac{\zeta_1}{\zeta_0}g^2\right)
+\frac{1}{2}\ln\det\left[\delta^{ab}\left(\delta^{\mu\nu}(\partial^2+m^2)
-\left(1-\frac{1}{\al}\right)\partial^\mu\partial^\nu\right)\right]\nonumber\\
&=&\frac{\s^2}{2\zeta_0}\left(1-\frac{\zeta_1}{\zeta_0}g^2\right)+\frac{N_A}{2}
\left[(d-1)\textrm{tr}\ln\left(\partial^2+m^2\right)+\textrm{tr}\ln\left(
\partial^2+\al m^2\right)\right]\;.
\end{eqnarray}
In dimensional regularization and using the
$\overline{\textrm{MS}}$ scheme, one finds
\begin{eqnarray}\label{rge27}
V_1(\sigma)&=&\frac{\s^2}{2\zeta_0}\left(1-\frac{\zeta_1}{\zeta_0}g^2\right)
+\frac{3N_A}{64\pi^2}\frac{g^2\s^2}{\zeta_0^2}
\left(-\frac{5}{6}+\ln\frac{g\s}{\zeta_0\omu^2}\right)\nonumber\\&+&
\frac{N_A}{64\pi^2}\frac{\al^2
g^2\s^2}{\zeta_0^2}\left(-\frac{3}{2}+\ln\frac{\al
g\s}{\zeta_0\omu^2}\right)\;,
\end{eqnarray}
where $\omu$ is the renormalization scale. It can be easily
checked that the infinities in the effective potential cancel when
the counterterms are included.

\noindent Next, we look for a non-trivial minimum of the effective
potential, which amounts to solving the gap equation
$\frac{dV}{d\s}=0$. To avoid possibly large logarithms, we will
set $\omu^2=m^2=\frac{g\s}{\zeta_0}$ in the gap equation,
\begin{eqnarray}\label{gapequation1}
    \left.\frac{dV}{d\sigma}\right|_{\omu^2=\frac{g\sigma}{\zeta_0}}&=&
    \frac{\sigma}{\zeta_0}\left(1-\frac{\zeta_1}{\zeta_0}g^2\right)
    +\frac{3N_A}{32\pi^2}\frac{g^2\sigma}{\zeta_0^2}\left(-\frac{5}{6}\right)+\frac{3N_A}{64\pi^2}\frac{g^2\sigma}{\zeta_0^2}\nonumber\\
    &+&\frac{N_A}{32\pi^2}\frac{\al^2g^2\sigma}{\zeta_0^2}\left(-\frac{3}{2}+\ln\al\right)+\frac{N_A}{64\pi^2}\frac{\al^2
    g^2\sigma}{\zeta_0^2}=0,
\end{eqnarray}
and use the RGE to sum leading logarithms. Defining
$y\equiv\frac{g^2N}{16\pi^2}$, we find as a solution of
(\ref{gapequation1})
\begin{equation}\label{rge28}
    \sigma=0 \textrm{ or
}y=\frac{C_A\zeta_0}{16\pi^2\zeta_1+\frac{N_A}{2}\left(1+\al^2-\al^2\ln
\left|\al\right|\right)}\;.
\end{equation}
The first solution corresponds to the trivial vacuum, while the
second one leads to
\begin{equation}\label{rge29}
    m=\lms e^{\frac{3}{22y}}\;,
\end{equation}
where the 1-loop formula for the coupling constant
\begin{equation}\label{rge30}
    g^{2}(\omu)=\frac{1}{\beta_0\ln\frac{\omu^2}{\lms^2}}\;,
\end{equation}
was used. The vacuum energy is given by
\begin{equation}\label{rge31}
    \Evac=-\frac{1}{2}\frac{N_A}{64\pi^2}\left(3+\al^2\right)m^4\;.
\end{equation}
We now consider the numerical evaluation of our results and
restrict ourselves to the colour group $SU(3)$. For $SU(N)$ one
has $T_F=\frac{1}{2}$, $C_F=\frac{N^2-1}{2N}$, $C_A=N$ and
$N_A=N^2-1$. For completeness, we quote the results for the Landau
gauge $\al=0$.
\begin{eqnarray}
\label{rge32a}y^{\textrm{\tiny{Landau}}}&=&\frac{36}{187}\approx0.193\; ,\\
\label{rge32b}m^{\textrm{\tiny{Landau}}}&=&e^{\frac{17}{24}}\lms\approx
2.031\lms\;,\\
\label{rge32c}\Evac^{\textrm{\tiny{Landau}}}&=&-\frac{3}{16\pi^2}
e^{\frac{17}{6}}\lms^4\approx-0.323\lms^4\;.
\end{eqnarray}
The results for general $\al$ are displayed in the Figures 1-3.
\begin{figure}[h]\label{fig1}
\begin{tabular}{ccc}
  \scalebox{0.6}{\includegraphics{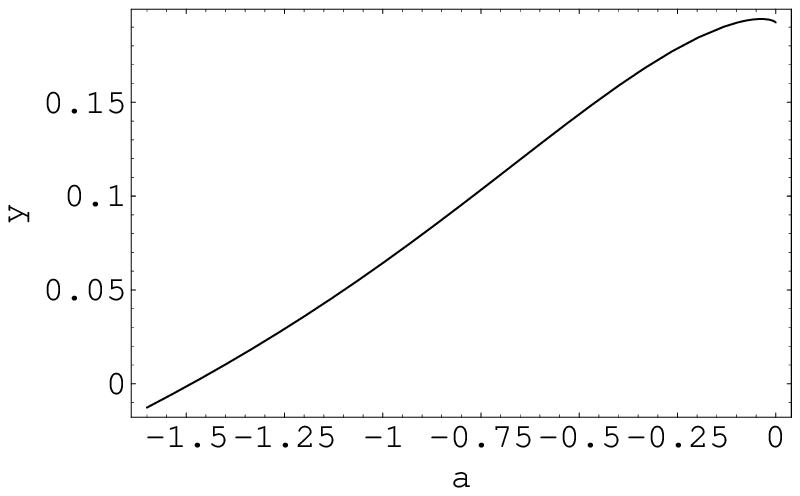}} &
\scalebox{0.6}{\includegraphics{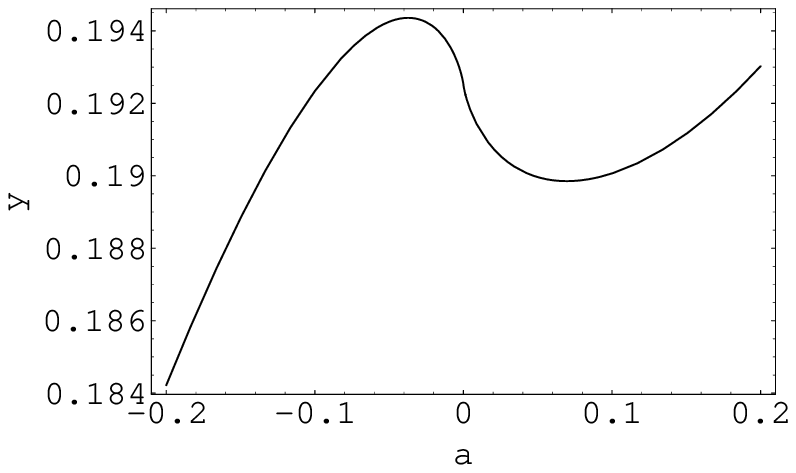}} &
\scalebox{0.6}{\includegraphics{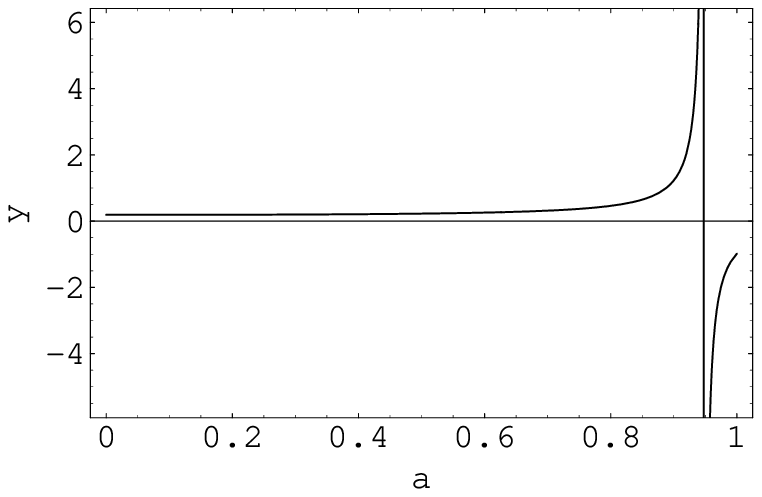}}\\
\end{tabular}
\caption{$y$ as a function of $\al$.}
\end{figure}
\begin{figure}[h]\label{fig2}
\begin{tabular}{ccc}
  \scalebox{0.6}{\includegraphics{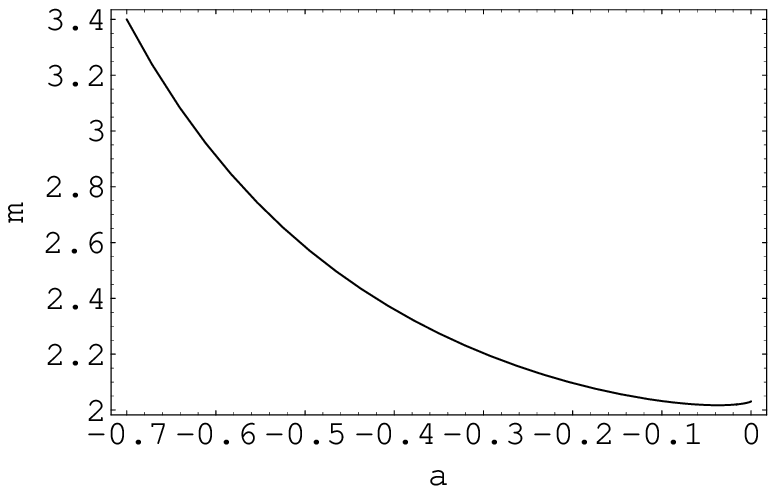}} &
\scalebox{0.6}{\includegraphics{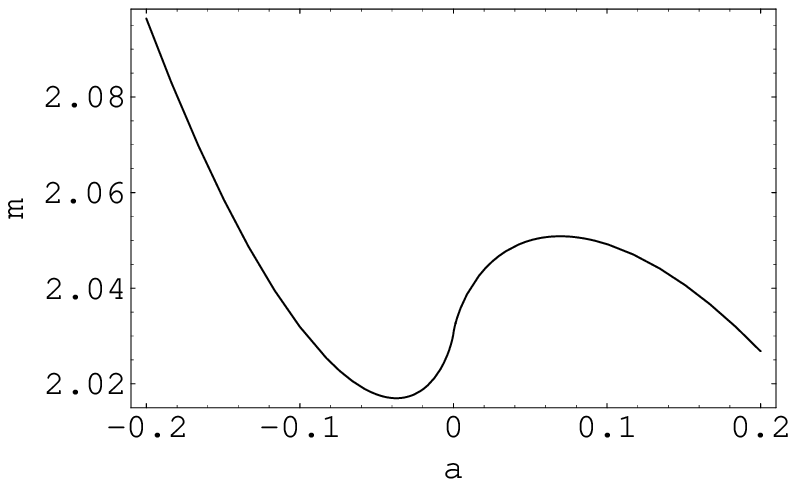}} &
\scalebox{0.6}{\includegraphics{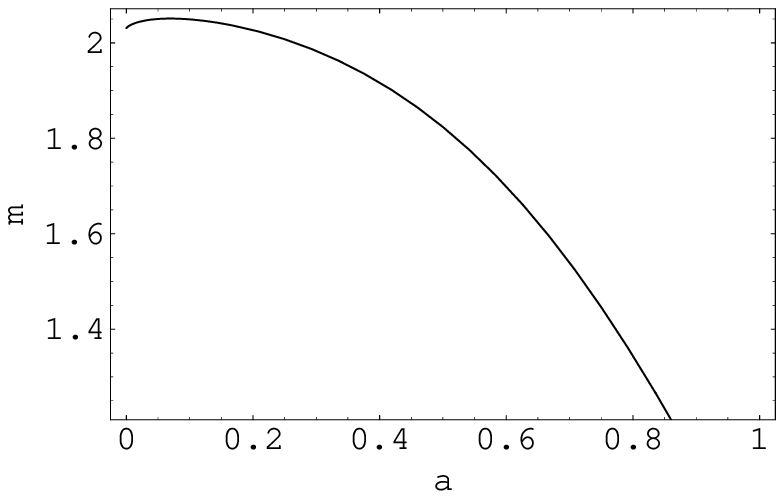}}\\
\end{tabular}
\caption{$m$ as a function of $\al$.}
\end{figure}
\begin{figure}[ht]
\begin{tabular}{ccc}
  \scalebox{0.6}{\includegraphics{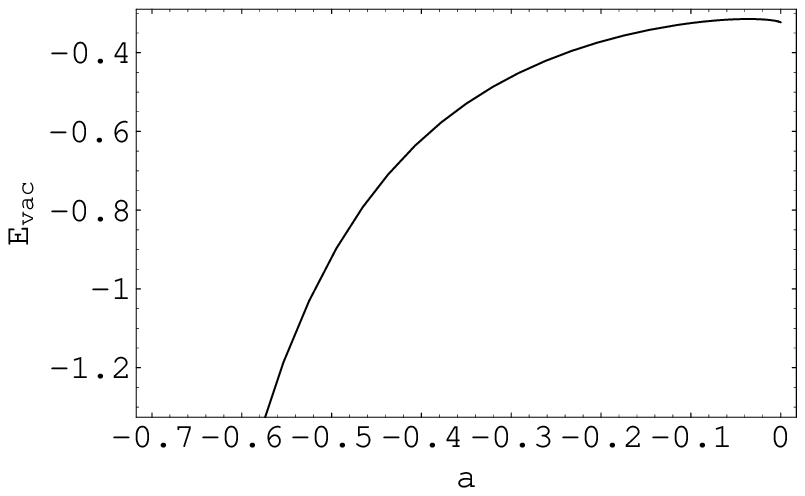}} &
\scalebox{0.6}{\includegraphics{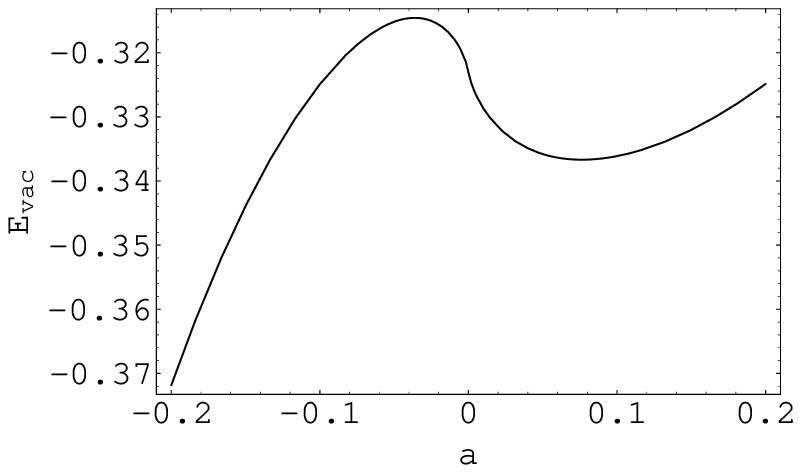}} &
\scalebox{0.6}{\includegraphics{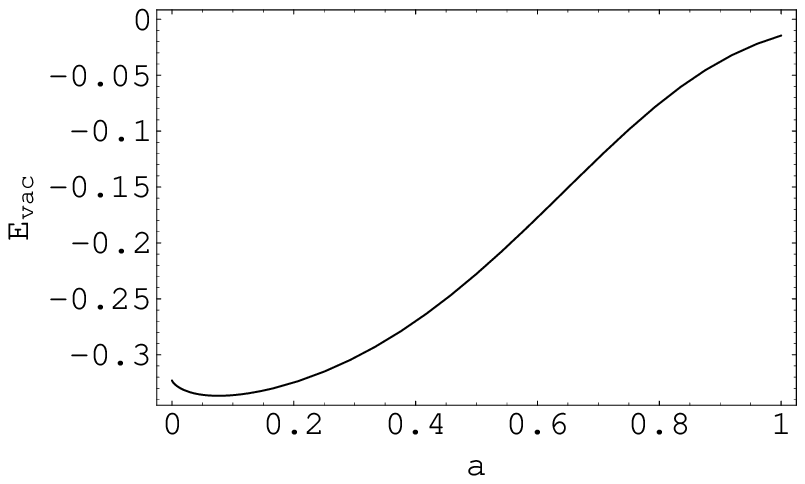}}\\
\end{tabular}
\caption{$\Evac$ as a function of $\al$.}
\end{figure}
For the moment, we have set $C_0=C_1=0$. Evidently, $y$ should
certainly be positive and also relatively small to have a sensible
expansion. Hence, we conclude from Figure 1 that we should
restrict the range of values for $\al$ further. We also see that
$m$ becomes rapidly larger and $\Evac$ becomes rapidly more and
more negative as $\alpha$ gets more negative. A more urgent
problem is the fact that the vacuum energy $\Evac$ depends on the
gauge parameter $\al$. Since $\Evac$ is a physical quantity, it
should be independent on the gauge parameter $\al$. In the next
section, we shall give a detailed account of this gauge parameter
dependence. We shall see that it is related to the impossibility
of evaluating the effective potential to arbitrary high loop
orders. Further, we shall provide a simple way to circumvent this
problem and obtain a vacuum energy which is independent of $\al$.

\section{Investigation of the gauge parameter dependence.}
One possible explanation as to why $\Evac$ depends on $\al$ could
reside in the values of the constants of integration $C_0$ and
$C_1$ we have chosen. With another choice for these constants, it
could be that $\Evac$ does not depend in $\alpha$, or equivalently
$\Evac=\Evac^{\textrm{\tiny{Landau}}}$. This can be investigated
by considering the expression (\ref{rge31}) for
$\Evac(\al,C_0,C_1)$. In order to have the same $\Evac$ for each
value of $\al$, we should solve the following equation
\begin{eqnarray}\label{rge33}
    \frac{d\Evac}{d\al}=0&\Leftrightarrow&2\al
    m^4+4(\al^2+3)m^3\frac{dm}{d\al}=0\nonumber\\
    &\Leftrightarrow&\al-\frac{3}{11y^2}(3+\al^2)\left(\frac{\partial
y}{\partial\al}+
    \frac{\partial y}{\partial \zeta_0}\frac{\partial\zeta_0}{\partial
\al}+\frac{\partial y}{\partial
\zeta_1}\frac{\partial\zeta_1}{\partial
    \al}\right)=0\;,
\end{eqnarray}
in terms of $C_0$ and $C_1$. However, the solutions of this
equation depend on $\alpha$, and this is not allowed since $C_0$
and $C_1$ should be $\alpha$ independent constants. This means
that the $\al$-dependence of $\Evac$ cannot be eliminated by a
suitable choice of $C_0$ and $C_1$.
\subsection{BRST symmetry and gauge parameter independence.}
Let us now turn to a more general analysis. Consider again the
generating functional (\ref{rge13}). We have the following
identification, ignoring the overall normalization factors
\begin{eqnarray}\label{rge34}
    \exp-i\mw(J)&=&\int [D\varphi]\exp iS_\sigma(J)\nonumber\\
    &=&\frac{1}{N}\int
    [D\varphi D\sigma]\exp i\left[S(J)+\int
d^{4}x\left(-\frac{1}{2\zeta}\left(%
\frac{\sigma}{g}-\frac{A_\mu^2}{2}-\zeta
J\right)^{2}\right)\right]\;,
\end{eqnarray}
where $S(J)$ and $S_\sigma(J)$ are given respectively by
(\ref{d2}), and (\ref{rge13bis}). Since
\begin{equation}
\frac{d}{d\alpha }\frac{1}{N}\int [D\sigma ]\exp \left[ i\int
d^{4}x\left( -%
\frac{1}{2\zeta }\left( \frac{\sigma }{g}-\frac{A_\mu^2}{2}-\zeta
J\right) ^{2}\right) \right] =\frac{d}{d\alpha }1=0\;,
\label{rge35}
\end{equation}
we find
\begin{equation}
-\frac{d\mathcal{W}(J)}{d\alpha }=\left\langle s\int d^{4}x\left(
\frac{%
\overline{c}b}{2}\right) \right\rangle _{J=0}+\textrm{terms
proportional to }J\;, \label{rge36}
\end{equation}
which follows by noticing that
\begin{eqnarray}
\frac{dS(J)}{d\alpha } &=&\int d^{4}x\left(
\frac{b^{a}b^{a}}{2}+\frac{
\partial \zeta }{\partial \alpha }\frac{J^{2}}{2}\right)   \nonumber  \\
&=&s\int d^{4}x\left( \frac{\overline{c}b}{2}\right)
+\textrm{terms proportional to }J\;.   \label{rge37}
\end{eqnarray}
We see that the first term in the right hand side of (\ref{rge37})
is an exact BRST variation. As such, its vacuum expectation value
vanishes. This is the usual argument to prove the gauge parameter
independence in the BRST framework \cite{book}. Of course, this is
based on the assumption that the BRST symmetry is not broken.
Notice therefore that there does not exist an operator
$\mathcal{G}$ with $A_\mu^2=s\mathcal{G}$, so that a non-vanishing
vacuum expectation value for the condensate $\left\langle
A_\mu^2\right\rangle$ does not break the BRST invariance. Indeed,
from
\begin{equation}\label{rge38}
    s\s=\frac{g}{2}sA_\mu^2=-gA_\mu^a \p^\mu c^a\;,
\end{equation}
one can easily check that
\begin{equation}\label{rge39}
    s\int d^{4}x\mathcal{L}(A_\mu,\s)=0\;,
\end{equation}
so that we have a BRST invariant $\sigma$-action.

\noindent The rest of the argument is based on the fact that $J=0$
when the vacuum is considered, so that we are left with only the
BRST exact term in (\ref{rge36}). More formally, the effective
action $\Gamma(\sigma)\equiv\Gamma\left(\frac{\s}{g}\right)$ is
related to $\mathcal{W}(J)$ through a Legendre transformation
\begin{equation}
\Gamma \left( \frac{\sigma }{g}\right) =-\mathcal{W}(J)-\int
d^{4}yJ(y)\frac{%
\sigma (y)}{g}\;.  \label{rge40}
\end{equation}
The effective potential $V(\sigma )$ is then defined as
\begin{equation}
-V(\sigma )\int d^{4}x=\Gamma \left( \frac{\sigma }{g}\right)\;.
\label{rge41}
\end{equation}
Let $\sigma _{min}$ be the solution of
\begin{equation}
\frac{dV(\sigma )}{d\sigma } =0\;. \label{rge42}
\end{equation}
Since
\begin{equation}\label{rge42bis}
    \frac{\delta }{\delta \left( \frac{\sigma }{g}\right) }\Gamma
    =-J\;,
\end{equation}
one finds
\begin{equation}
\sigma =\sigma _{min}\Rightarrow J=0\;,  \label{rge43}
\end{equation}
and invoking (\ref{rge43}), from (\ref{rge40}) and (\ref{rge41})
we derive
\begin{equation}
\left. \frac{d}{d\alpha }V(\sigma )\right| _{\sigma =\sigma
_{min}}\int d^{4}x=\left. \frac{d}{d\alpha }\mathcal{W}(J)\right|
_{J=0}\;.  \label{rge44}
\end{equation}
Finally, combining (\ref{rge36}) and (\ref{rge44})
\begin{equation}
\left. \frac{d}{d\alpha }V(\sigma )\right| _{\sigma =\sigma
_{min}}=0\;. \label{rge45}
\end{equation}
{}From this, we conclude that the vacuum energy $\Evac$ should be
independent of the gauge parameter $\al$.

\noindent Apparently, our explicit result (\ref{rge31}) for
$\Evac$ is not in agreement with the above proof that $\Evac$ is
the same for each $\al$. If we examine the proof in more detail we
notice that a key argument is that $J$ becomes zero at the end of
the calculation. In practice, this is achieved by solving the gap
equation. Now, in a power series expansion in the coupling
constant, the derivative of the effective potential with respect
to $\sigma$ is something of the form
\begin{equation}\label{rge46}
    \left(v_0+v_1g^2+0(g^4)\right)\sigma\;,
\end{equation}
where we assume that we work up to order $g^2$ and that we have
chosen $\omu$ so that the logarithms vanish. Then, the gap
equation corresponding to (\ref{rge46}) reads
\begin{equation}\label{rge47}
v_0+v_1g^2+O(g^4)=0\;.
\end{equation}
Due to (\ref{rge41}) and (\ref{rge42bis}), one also has
\begin{equation}\label{rge48}
J=g\left(v_0+v_1g^2+O(g^4)\right)\sigma\;.
\end{equation}
This means that, if we solve the gap equation (\ref{rge47}) up to
certain order, we have
\begin{equation}\label{rge49}
J=g\left(0+O(g^4)\right)\sigma\;.
\end{equation}
We also have
\begin{eqnarray}
\frac{\p\zeta}{\p\al}=\frac{\p\zeta_0}{\p\al}\frac{1}{g^2}+\frac{\p\zeta_1}
{\p\al}+O(g^4)\;.
\end{eqnarray}
So, working to the order we are considering
\begin{equation}\label{rge50}
\frac{\p\zeta}{\p\al}J^2=\left(\frac{\p\zeta_0}{\p\al}v_0^2+\left(
\frac{\p\zeta_0}{\p\al}2v_0v_1+\frac{\p\zeta_1}{\p\al}v_0^2\right)g^2+O(g^4)
\right)\sigma^2\;.
\end{equation}
{}From the square of the gap equation (\ref{rge47}),
\begin{equation}\label{rge51}
v_0^2+2v_1v_0g^2+O(g^4)=0\;,
\end{equation}
it follows that
\begin{equation}\label{rge52}
\frac{\p\zeta}{\p\al}J^2=\left(\frac{\p\zeta_1}{\p\al}v_0^2g^2+O(g^4)\right)
\sigma^2\;.
\end{equation}
We see that, if one consistently works to the order we are
considering, terms such as $\frac{\p\zeta}{\p\al}J^2$ do not equal
zero although $J=0$ to that order. Terms like those on the right
hand side of (\ref{rge52}) are cancelled by terms which are
formally of higher order. This has its consequences for the terms
proportional to $J$ in (\ref{rge36}). If one were able to work to
infinite order, the problem would not arise. However, we do not
have this ability, and we are faced with a gauge parameter
dependence slipping into $\Evac$.

\subsection{Circumventing the gauge parameter dependence.}
We could resolve this issue by saying that the gauge parameter
dependence of the vacuum energy should become less and less severe
as we go to higher orders, and that eventually it will drop out if
we go to infinite order. However, this is not very satisfactory,
especially since we can surely never calculate the potential up to
infinite order. Also as is clear from the quite complicated
expression for $\zeta_1(\al)$, which will enter the differential
equation for $\zeta_2(\al)$, a 2-loop evaluation of the effective
potential is already out of the question.

\noindent Therefore, we could try to modify the LCO formalism in
order to circumvent the gauge parameter dependence of $\Evac$.
Therefore, we consider the following action
\begin{eqnarray}
\widetilde{S}(\wj)&=&S_{YM}+S_{GF+FP}+\int d^{4}x\left[
\widetilde{J}\mf(\al)\frac{A_\mu^2}{2} +\frac{\zeta }{2}%
\mf^2(\al)\widetilde{J}^{2}\right]\;, \label{rge53}
\end{eqnarray}
instead of (\ref{d2}) where, for the moment, $\mf(\al)$ is an
arbitrary function of $\al$ of the form
\begin{equation}\label{rge54}
    \mf(\al)=1+f_{0}(\al)g^2+f_1(\al)g^4+O(g^6)\;,
\end{equation}
and $\wj$ is now the source. The generating functional becomes
\begin{equation}\label{rge55}
    \exp-i\mathcal{\widetilde{W}}(\wj)=\int[D\phi]\exp
i\widetilde{S}(\wj)\;.
\end{equation}
Taking the functional derivative of $\mathcal{\widetilde{W}}(\wj)$
with respect to $\wj$, we obtain
\begin{equation}
\left. \frac{\delta \mathcal{\widetilde{W}}(\wj)}{\delta
\wj}\right|
_{\wj=0}=-\mf(\al)\left\langle\frac{A_\mu^2}{2}\right\rangle\;.
\label{rge56}
\end{equation}
Again, we insert unity via
\begin{equation}  \label{rge57}
1=\frac{1}{N}\int [D\wsigma]\exp\left[i\int
d^{4}x\left(-\frac{1}{2\zeta}\left(%
\frac{\wsigma}{g\mf(\al)}-\frac{A_\mu^2}{2}-\zeta \wj\mf(\al)
\right)^{2}\right)\right]\;,
\end{equation}
to arrive at the following renormalized Lagrangian
\begin{eqnarray}\label{rge58}
\mathcal{\widetilde{L}}(A_\mu,\wsigma)=-\frac{1}{4}F_{\mu\nu}^{2}
+\mathcal{L}_{GF+FP}-\frac{\wsigma^2}{2g^2\mf^2(\al)Z_\zeta\zeta}
    +\frac{1}{2}\frac{Z_2}{g^2\mf(\al)Z_\zeta\zeta}g\wsigma
A_{\mu}^2-\frac{1}{8}\frac{Z_{2}^2}{Z_\zeta\zeta}\left(A_\mu^2\right)^2
+\wj\frac{\wsigma}{g}\;.
\end{eqnarray}
{}From the generating functional
\begin{equation}\label{rge59}
    \exp-i\mathcal{\widetilde{W}}(\wj)=\int[D\phi]\exp i\int
    d^{4}x\mathcal{\widetilde{L}}(A_\mu,\wsigma)\;,
\end{equation}
it follows that
\begin{equation}
\left. \frac{\delta \mathcal{\widetilde{W}}(\wj)}{\delta
\wj}\right|
_{\wj=0}=-\left\langle\frac{\wsigma}{g}\right\rangle\Rightarrow\left\langle
\wsigma\right\rangle
=g\mf(\al)\left\langle\frac{A_\mu^2}{2}\right\rangle\;,
\label{rge60}
\end{equation}
where the anomalous dimension of $\wsigma$ equals
\begin{equation}\label{rge61}
    \gamma_{\wsigma}(g^2)=\frac{\mu}{\wsigma}\frac{\p \wsigma}{\p
\mu}=\frac{\beta(g^2)}{2g^2}+\gamma_{A^2}(g^2)+\mu\frac{\p\ln\mf(\al)}
{\p\mu}\;.
\end{equation}
The lowest order gluon mass is now provided by
\begin{equation}\label{rge61bis}
    m^2=\frac{g\wsigma}{\zeta_0}\;,
\end{equation}
and the vacuum configurations are now determined by solving
\begin{equation}\label{rge62}
    \frac{d\widetilde{V}(\wsigma)}{d\wsigma}=0\;.
\end{equation}
with $\widetilde{V}(\wsigma)$ the effective potential. In the
$\MSbar$ scheme, the 1-loop effective potential reads
\begin{eqnarray}\label{rge63}
\widetilde{V}_1(\wsigma)&=&\frac{\wsigma^2}{2\zeta_0}\left(1-\left(2f_0
+\frac{\zeta_1}{\zeta_0}\right)g^2
+\frac{2}{\varepsilon}\frac{N_A}{16\pi^2}\frac{g^2}{\zeta_0}\left(\frac{3}{2}
+\frac{\al^2}{2}\right)\right)\nonumber\\
&+&\frac{3\left(N_A\right)}{64\pi^2}\frac{g^2\wsigma^2}{\zeta_0^2}
\left(-\frac{2}{\varepsilon}-\frac{5}{6}+\ln\frac{g\wsigma}{\zeta_0\omu^2
}\right)+\frac{N_A}{64\pi^2}\frac{\al^2
g^2\wsigma^2}{\zeta_0^2}\left(-\frac{2}{\varepsilon}-\frac{3}{2}+\ln\frac
{\al g\wsigma}{\zeta_0\omu^2}\right)
\nonumber\\
&=&\frac{\wsigma^2}{2\zeta_0}\left(1-\left(2f_0+\frac{\zeta_1}{\zeta_0}
\right)g^2\right)
+\frac{3N_A}{64\pi^2}\frac{g^2\wsigma^2}{\zeta_0^2}
\left(-\frac{5}{6}+\ln\frac{g\wsigma}{\zeta_0\omu^2}\right)\nonumber\\&+&
\frac{N_A}{64\pi^2}\frac{\al^2
g^2\wsigma^2}{\zeta_0^2}\left(-\frac{3}{2}+\ln\frac{\al
g\wsigma}{\zeta_0\omu^2}\right)\;.
\end{eqnarray}
We included the counterterm contribution here to illustrate
explicitly that $\widetilde{V}_1(\wsigma)$ is finite. With
(\ref{rge61}), it can also be checked that
\begin{equation}\label{rge64}
    \mu\frac{d}{d\mu}\widetilde{V}_1(\wsigma)=0+O(g^4)\;.
\end{equation}
Now, we can continue with the determination of the 1-loop vacuum
energy, which will not only depend on $\al$, $C_0$ and $C_1$, but
also on $f_0(\al)$. We will determine an expression for $f_0(\al)$
so that $\Evac(\al,C_0,C_1,f_0(\al))$ does not depend on $\al$. In
the meantime, we could also absorb the constants of integration
$C_0$ and $C_1$ in $f_0(\al)$ so that $\Evac$ does not depend on
them either. Based on this, we will immediately set $C_0=C_1=0$.
As usual, we put $\omu^2=\frac{g\wsigma}{\zeta_0}$ in the gap
equation, which now reads
\begin{eqnarray}\label{gapequation2}
    \left.\frac{d\widetilde{V}}{d\wsigma}\right|_{\omu^2=\frac{g\wsigma}{\zeta_0}}&=&
    \frac{\wsigma}{\zeta_0}\left(1-\left(2f_0+\frac{\zeta_1}{\zeta_0}\right)g^2\right)
    +\frac{3N_A}{32\pi^2}\frac{g^2\wsigma}{\zeta_0^2}\left(-\frac{5}{6}\right)+\frac{3N_A}{64\pi^2}\frac{g^2\wsigma}{\zeta_0^2}\nonumber\\
    &+&\frac{N_A}{32\pi^2}\frac{\al^2g^2\wsigma}{\zeta_0^2}\left(-\frac{3}{2}+\ln\al\right)+\frac{N_A}{64\pi^2}\frac{\al^2
    g^2\wsigma}{\zeta_0^2}=0,
\end{eqnarray}
and use the RGE to sum the leading logarithms. One finds, in
addition to the trivial solution $\wsigma=0$,
\begin{eqnarray}
\label{rge65a}y&=&\frac{C_A\zeta_0}{16\pi^2\left(2f_0\zeta_0+\zeta_1\right)
+\frac{N_A}{2}\left(1+\al^2-\al^2\ln\left|\al\right|\right)}\;,\\
    \label{rge65b}m&=&\lms e^{\frac{3}{22y}}\;,\\
\label{rge65c}\Evac&=&-\frac{1}{2}\frac{N_A}{64\pi^2}\left(3+\al^2\right)
m^4\;.
\end{eqnarray}
In principle, the analytic solution for $f_0(\al)$ can be obtained
by solving the following differential equation
\begin{eqnarray}\label{rge66}
    \frac{d\Evac}{d\al}=0&\Leftrightarrow&2\al
    m^4+4(\al^2+3)m^3\frac{dm}{d\al}=0\nonumber\\
    &\Leftrightarrow&\al-\frac{3}{11y^2}(3+\al^2)\left(\frac{\partial
y}{\partial\al}+
    \frac{\partial y}{\partial \zeta_0}\frac{\partial\zeta_0}{\partial \al}
    +\frac{\partial y}{\partial \zeta_1}\frac{\partial\zeta_1}{\partial
\al}+\frac{\partial y}{\partial f_0}\frac{\partial f_0}{\partial
    \al}\right)=0\;.
\end{eqnarray}
The quantity $f_0(\al)$ constructed in this fashion will ensure
$\Evac(\al)$ is independent of the gauge parameter $\al$. However,
we still have the freedom of choosing an initial condition. We
will determine $f_0(\al)$ so that
$\Evac(\al)=\Evac(0)\equiv\Evac^{\textrm{\tiny{Landau}}}$. This
amounts to choosing $f_0(\al=0)=0$. We can justify this choice
based on our remark in the introduction, which is that $A_\mu^2$
coincides with the gauge invariant quantity $A^2_{\min}$ in the
Landau gauge in the FMR. Since our calculation is based on a
perturbative expansion around $A_\mu^a=0$, which lies within the
FMR, we stay within the FMR
\cite{semenov,Zwanziger:tn,Stodolsky:2002st}.

\noindent Unfortunately, the differential equation (\ref{rge66})
is very hard to solve analytically. We could solve (\ref{rge66})
and consequently $y$, $m$ and $\Evac$ numerically. However, there
is a more elegant way to obtain the analytical solution for
$f_0(\al)$. Considering the colour group $SU(3)$ for simplicity,
then since we know that by construction that
$\Evac=\Evac^{\textrm{\tiny{Landau}}}$, we are able to write down
the analytical solution for $m$ as
\begin{equation}\label{rge67}
    m=\left(\frac{3e^{17/6}}{3+\al^2}\right)^{1/4}\lms\;,
\end{equation}
where use was made of (\ref{rge32c}) and (\ref{rge65c}). Putting
(\ref{rge67}) in (\ref{rge65b}), we deduce that
\begin{eqnarray}\label{rge68}
    y=\frac{36}{66\ln\frac{3}{3+\al^2}+187}\;.
\end{eqnarray}
Combining (\ref{rge65a}) and (\ref{rge68}) finally gives the
analytic expression for $f_{0}(\al)$
\begin{equation}\label{rge69}
f_0(\al)=\frac{\frac{\zeta_0}{12}\left(66\ln\frac{3}{3+\al^2}+187\right)-
4\left(1+\al^2-\al^2\ln|\al|\right)-16\pi^2\zeta_1}{32\pi^2\zeta_0}\;.
\end{equation}
We have displayed $f_0(\al)$, $y(\al)$ and $m(\al)$ for the range
of values $-\frac{13}{3}$~$<$~$\al$~$<$~$\frac{13}{3}$ in Figures
4-6. As a check, we have also plotted, in Figure 7,
$\Evac(\al,f_0(\al))$ as given in (\ref{rge65c}) to verify that
$\Evac(\al,f_0(\al))=\Evac^{\textrm{\tiny{Landau}}}$. We observe
several features. Firstly, although $f_0(\al)$ has some
singularities in $\left]-\frac{13}{3},\frac{13}{3}\right[$, the
quantities $y$, $m$ and $\Evac$ are completely regular functions
of $\al$. Secondly, the expansion parameter $y$ remains relatively
small, which makes our numerical predictions at least
qualitatively trustworthy. Thirdly, we also see that the value for
the tree level mass does not change spectacularly in the
considered region. In the Feynman gauge $\al=1$, we have
$m^{\textrm{\tiny{Feynman}}}=1.89\lms$.
\begin{figure}[ht]\label{fig10}
    \begin{center}
        \scalebox{1}{\includegraphics{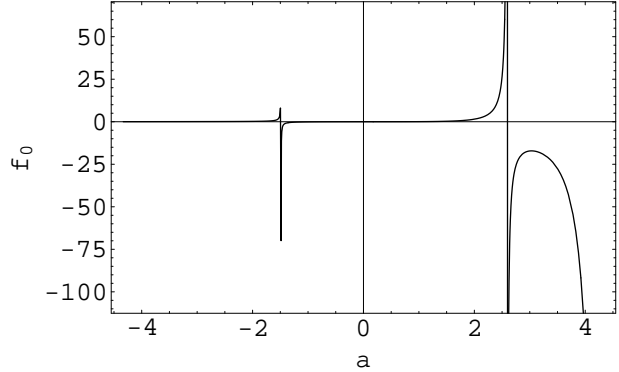}}
        \caption{$f_0$ as a function of $\al$ with
$-\frac{13}{3}<\al<\frac{13}{3}$.}
    \end{center}
\end{figure}
\begin{figure}[ht]\label{fig11}
    \begin{center}
        \scalebox{1}{\includegraphics{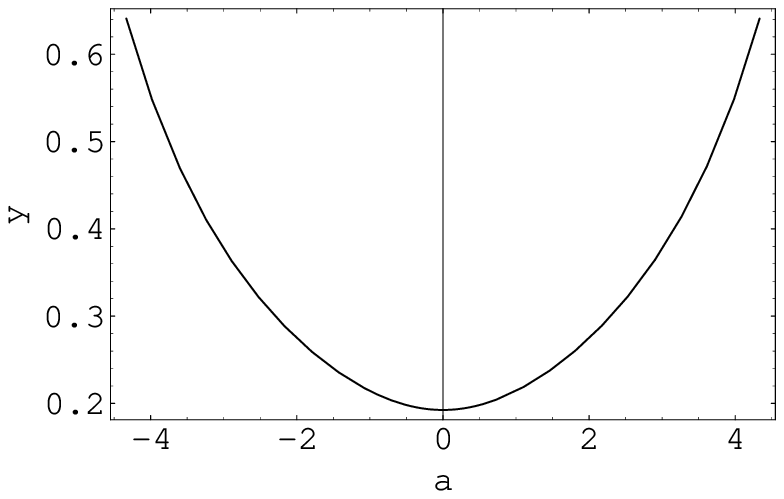}}
        \caption{$y$ as a function of $\al$ with
$-\frac{13}{3}<\al<\frac{13}{3}$.}
    \end{center}
\end{figure}
\begin{figure}[ht]\label{fig12}
    \begin{center}
        \scalebox{1}{\includegraphics{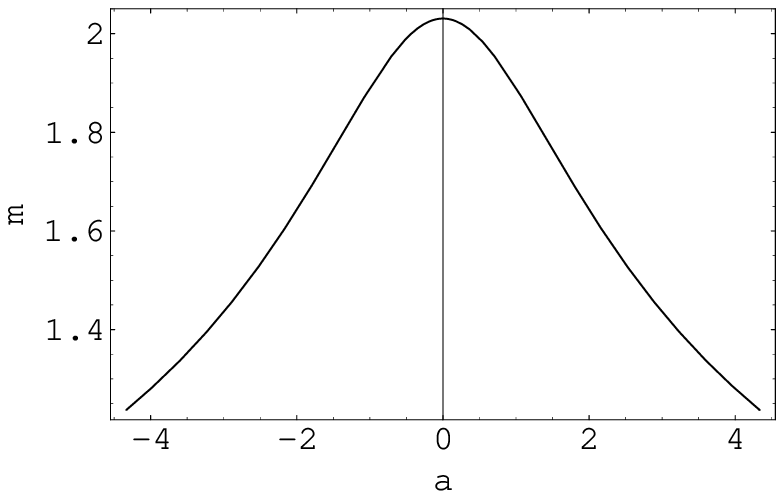}}
        \caption{$m$ as a function of $\al$ with
$-\frac{13}{3}<\al<\frac{13}{3}$.}
    \end{center}
\end{figure}
\begin{figure}[ht]\label{fig13}
    \begin{center}
        \scalebox{1}{\includegraphics{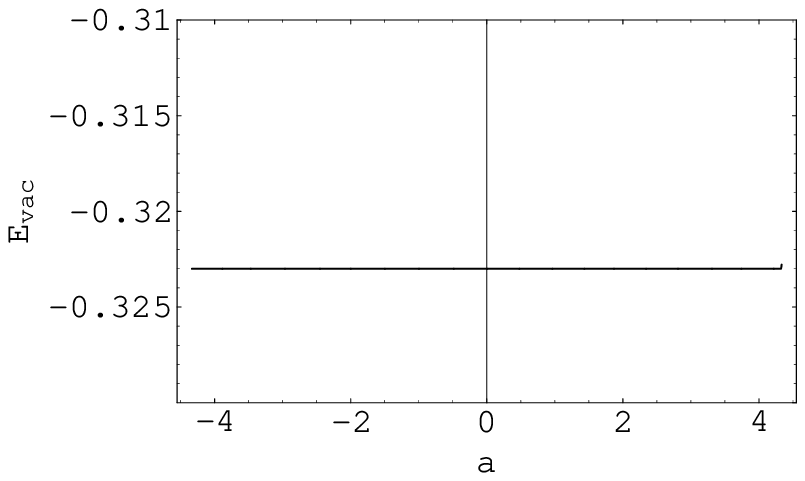}}
        \caption{$\Evac$ as a function of $\al$ with
$-\frac{13}{3}<\al<\frac{13}{3}$.}
    \end{center}
\end{figure}

\noindent Before ending this section, there are several other
points. We have determined $\mathcal{F}(\al)$ with the
renormalization scale $\omu$ chosen in such a way that the
logarithms vanish. Other choices of $\omu$ are of course also
valid. We did not explicitly write this $\omu$ dependence of
$\mathcal{F}(\al)$ in (\ref{rge54}).

\noindent Also, the procedure we have described here applies of
course at higher order. For example, at 2-loops, $f_1(\al)$ will
be required to remove the $\al$ dependence. If we were to work to
infinite order in $g^2$, we could transform the action
$\widetilde{S}(\wj)$ (\ref{rge53}) \emph{exactly} into the action
$S(J)$ (\ref{d2}) by means of the transformation
\begin{equation}\label{rge70}
    \wj=\frac{J}{\mf(\al)}\;.
\end{equation}
The corresponding transformation for the $\sigma$ and $\wsigma$
fields reads
\begin{equation}\label{rge71}
    \wsigma=\mf(\al)\sigma\;,
\end{equation}
which will transform the effective potential
$\widetilde{V}_{\infty}(\wsigma)$ \emph{exactly} into
$V_{\infty}(\sigma)$. As such, the constructed vacuum energy will
be the same in both cases and independent of the choice of $\al$.

\section{Gluon propagator in linear covariant gauges.}
In \cite{Langfeld:2001cz}, the gluon propagator in the Landau was
investigated, and a fit of the lattice results gave evidence for a
gluon mass. In the Landau gauge, the lattice also gives evidence
for the existence of a non-zero $\left\langle
A_\mu^2\right\rangle$ condensate, based on the discrepancy in the
10 GeV region, between the behaviour of the observed lattice gluon
propagator and strong coupling constant and the expected
perturbative behaviour. The results could be matched together
using an operator product expansion analysis with a non-zero
$\left\langle A_\mu^2\right\rangle$ condensate
\cite{Boucaud:2000nd,Boucaud:2001st,Boucaud:2002nc}. A combined
lattice fit resulted in $\left\langle A_\mu^2
\right\rangle_{\textrm{\tiny{OPE}}}\approx (1.64\textrm{GeV})^2$.
This quantity was obtained at a scale of 10 GeV in the MOM
renormalization scheme. Later, it was argued that this
$\left\langle A_\mu^2 \right\rangle_{\textrm{\tiny{OPE}}}$
condensate could be explained with instantons
\cite{Boucaud:2002nc}.

\noindent One will notice that we did not give the estimate for
$\left\langle A^2 \right\rangle$ itself. From the identification
(\ref{rge14}) and using the relation (\ref{rge25}) and the
explicit result (\ref{rge32b}), one finds
\begin{equation}\label{invoegsel}
\left\langle A_\mu^2
\right\rangle=-\frac{187}{52\pi^2}e^{\frac{17}{12}}\lms^2\approx-\left(0.29\textrm{GeV}\right)^2
\end{equation}
The extra minus sign arises because we have rotated from
Minkowskian to Euclidean space time to make possible a comparison
with the lattice. We used $\lms=0.233\textrm{GeV}$, which was the
value obtained in \cite{Boucaud:2001st}. We should be careful not
to misinterpret the relatively big difference between
$\left\langle A_\mu^2 \right\rangle_{\textrm{\tiny{OPE}}}$ and
(\ref{invoegsel}). Although our result is non-perturbative in
nature, it is still obtained in perturbation theory and as such it
only gives information from the high energy region (or short
range), while the OPE approach of Boucaud {\it et al} can only
describe the low energy (or long range) content of $\left\langle
A_\mu^2 \right\rangle$. It was already argued in
\cite{Gubarev:2000nz} that $\left\langle A_\mu^{2}\right\rangle$
can receive long and short range contributions. The minus sign in
front of our result has to do with the regularization and
renormalization of the quantity $\left\langle
A_\mu^{2}\right\rangle$. We refer to \cite{Verschelde:2001ia} for
more details.

\noindent To our knowledge, there has been little attention on the
lattice to the gluon propagator in a general linear covariant
gauge. Giusti {\it et al} managed to put the linear covariant
gauge on the lattice
\cite{Giusti:1996kf,Giusti:1999im,Giusti:2000yc,Giusti:2001kr}.
The tree level gluon propagator of Euclidean Yang-Mills theory
with a linear covariant gauge fixing is given by
\begin{equation}\label{prop1}
D_{\mu\nu}(q)=\frac{1}{q^2}\left(\delta_{\mu\nu}-(1-\alpha)\frac{q_\mu
    q_\nu}{q^2}\right)\;.
\end{equation}
This can be decomposed into the transverse and longitudinal parts
as
\begin{equation}\label{prop2}
D_{\mu\nu}(q)=\frac{1}{q^2}\left(\delta_{\mu\nu}-\frac{q_\mu
q_\nu}{q^2}\right)D^{T}(q)+\frac{q_\mu
q_\nu}{q^2}\frac{D^L(q)}{q^2}\;,
\end{equation}
where $D^T(q^2)$ is $q^2$ times the one used in
\cite{Giusti:1996kf,Giusti:1999im,Giusti:2000yc,Giusti:2001kr}. In
general, one determines $D^L(q)$ via the projector
\begin{eqnarray}
P^L_{\mu\nu}(q)= q_\mu q_\nu\;.
\end{eqnarray}
If there is a tree level gluon mass $m$ present, as in
(\ref{rge58}), the Euclidean gluon propagator in linear covariant
gauges reads
\begin{equation}\label{prop3}
D_{\mu\nu}(q)=\frac{1}{q^2+m^2}\left(\delta_{\mu\nu}-(1-\alpha)\frac{q_\mu
q_\nu}{q^2+\alpha
    m^2}\right)\;,
\end{equation}
with the value of $m$ given in (\ref{rge67}). The longitudinal
part of this propagator is
\begin{eqnarray}\label{prop4}
    D^L(q)=P^{L}_{\mu\nu}(q)D_{\mu\nu}(q)
        =\frac{1}{q^2+m^2}\left(q^2-(1-\alpha)\frac{q^4}{q^2+\alpha
        m^2}\right)\;.
\end{eqnarray}
$D^L(q)$ is plotted in Figure 8, again using
$\lms=0.233\textrm{GeV}$.
\begin{figure}[ht]\label{fig15}
    \begin{center}
        \scalebox{1.5}{\includegraphics{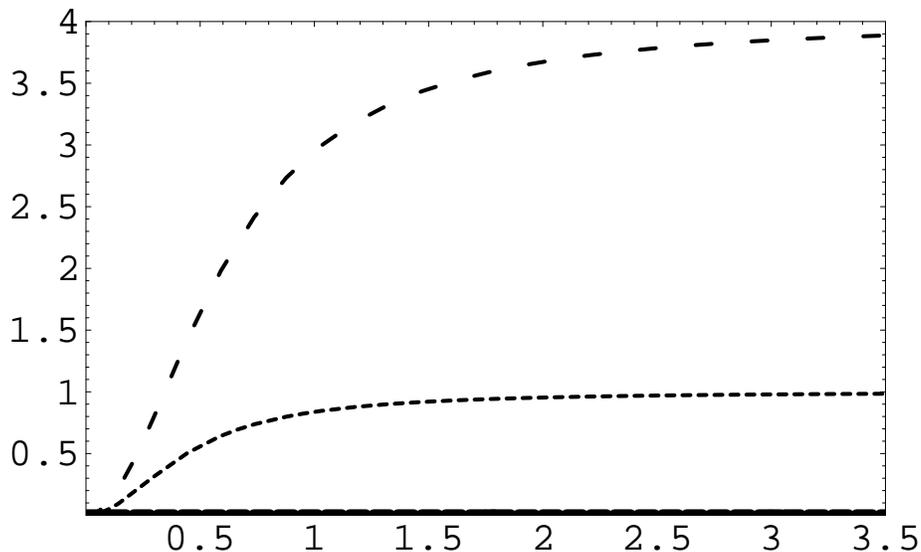}}
        \caption{$D_L(q)$ as a function of $q$ with
$0<q<3.5\textrm{GeV}$ for $\al=0$ (fat), $\al=1$ (dashed) and
$\al=4$ (wide dashed).}
    \end{center}
\end{figure}
Of course, we should not attach any firm meaning to this plot,
since we are only considering the tree level propagator and do not
include any renormalization effects. If we could calculate the
form factors, we would also inevitably encounter the problem of a
diverging perturbation theory in the infrared region. We cannot
make any conclusion about the behaviour of the propagator in the
IR from the above. Many other (non-perturbative) effects can
influence the propagators form in the IR. Nevertheless, it might
be worth noticing that the longitudinal part $D^L(q)$ is not
proportional to the gauge parameter. A similar behaviour was found
by Giusti \emph{et al}, see e.g. Figure 4 of \cite{Giusti:2000yc}.
This is already different from the perturbative prediction of
massless Yang-Mills theory with a linear covariant gauge fixing
\cite{Alkofer:2000wg}.

\section{Conclusion.}
We have considered Yang-Mills theories in linear covariant gauges
and constructed a renormalizable effective potential by means of
the local composite operator formalism for $A_\mu^2$. The
formation of the gluon condensate of mass dimension two is
favoured since it lowers the vacuum energy. As a result, the
gluons acquire a dynamical mass $m$. We discussed the gauge
parameter dependence of the resultant vacuum energy and observed
that this is due to the fact that we do not work up to infinite
order precision, but have to truncate the perturbative expansion
at a finite order. We explained how this gauge parameter
dependence can be avoided by a modification of our method.

\noindent Although there is limited lattice data available for the
general linear covariant gauges compared with the Landau gauge, it
would be interesting to calculate the form factor of the
longitudinal and transverse part of the gluon propagator to make a
more detailed comparison possible with the lattice results of
\cite{Giusti:1996kf,Giusti:1999im,Giusti:2000yc,Giusti:2001kr}. It
would also be useful to have direct evidence from the lattice
community that the $\left\langle A_\mu^2\right\rangle$ condensate
exists and that the gluons become massive, in analogy with the
Landau gauge. A further point worth investigating is the possible
existence of ghost condensates in the linear covariant gauges, as
is the case in the Landau gauge \cite{Lemes:2002rc,Dudal:2003dp}.
These condensates can modify the gluon propagator further.

\section*{Acknowledgments.}
D.~D. would like to thank K. Van Acoleyen and M. Picariello for
useful discussions. The Conselho Nacional de Desenvolvimento
Cient\'{\i}fico e Tecnol\'{o}gico (CNPq-Brazil), the SR2-UERJ and
the Coordena{\c{c}}{\~{a}}o de
Aperfei{\c{c}}%
oamento de Pessoal de N{\'\i}vel Superior (CAPES) are gratefully
acknowledged for financial support.

\end{document}